\documentclass[a4paper,fleqn,usenatbib,useAMS]{mnras}

\usepackage{graphicx}	
\usepackage{amsmath}	
\usepackage{amssymb}	
\usepackage{multicol}        
\usepackage{bm}		
\usepackage{pdflscape}	

\usepackage[T1]{fontenc}
\usepackage{ae,aecompl}

\title[J021659-044920: a relic GRG]{J021659-044920: a relic giant radio galaxy at $z$ $\sim$ 1.3}

\author[P. Tamhane et al.]{P. Tamhane,$^{1,2}$\thanks{E-mail:
		\href{mailto:ptamhane@students.iiserpune.ac.in}{ptamhane@students.iiserpune.ac.in}}
		Y. Wadadekar,$^2$\thanks{E-mail: \href{mailto:yogesh@ncra.tifr.res.in}
		{yogesh@ncra.tifr.res.in}}
		A. Basu,$^3$
		V. Singh,$^4$
		C. H. Ishwara-Chandra,$^2$
\newauthor 
        A. Beelen$^5$,
and S. Sirothia$^{2,6,7}$ \\
\\
$^{1}$Indian Institute of Science Education and Research, Dr Homi Bhabha Road, Pashan, Pune 411008, India\\
$^2$National Centre for Radio Astrophysics, TIFR, Post Bag 3, Ganeshkhind, Pune 411007, India \\
$^3$Max-Planck-Institut f\"{u}r Radioastronomie, Auf dem H{\"u}gel 69, D-53121 Bonn, Germany\\
$^4$Astrophysics and Cosmology Research Unit, School of Chemistry and Physics,
University of KwaZulu-Natal, Durban 4000, South Africa\\
$^5$Institut d'Astrophysique Spatiale, B$\hat{\rm a}$t. 121, Universit{\'e} Paris-Sud, F-91405 Orsay Cedex, France\\
$^6$Square Kilometre Array South Africa, 3rd Floor, The Park, Park Road, Pinelands, 7405, South Africa\\
$^7$Department of Physics and Electronics, Rhodes University, PO Box 94, Grahamstown 6140, South Africa}

\date{Published in MNRAS; volume 453, issue 3, pages 2438-2446}

\pubyear{2015}

\begin{document}
\label{firstpage}
\pagerange{\pageref{firstpage}--\pageref{lastpage}}
\maketitle

\begin{abstract}

We report the discovery of a relic Giant Radio Galaxy (GRG) J021659-044920 at
redshift $z \sim 1.3$ that exhibits large-scale extended, nearly co-spatial,
radio and X-ray emission from radio lobes, but no detection of Active Galactic Nuclei core, jets
and hotspots. The total angular extent of the GRG at the observed frame 0.325
GHz, using Giant Metrewave Radio Telescope observations is found to be ${\sim}$ 2.4 arcmin, that
corresponds to a total projected linear size of $\sim$ 1.2 Mpc. The integrated
radio spectrum between 0.240 and 1.4 GHz shows high spectral curvature
(${\alpha}_{\rm 0.610~GHz}^{\rm 1.4~GHz} - {\alpha}_{\rm 0.240~GHz}^{\rm
0.325~GHz}$ $>$ 1.19) with sharp steepening above 0.325 GHz, consistent with
relic radio emission that is $\sim$ 8 $\times$ 10$^{6}$ yr old.  The radio
spectral index map between observed frame 0.325 and 1.4~GHz for the two lobes
varies from 1.4 to 2.5 with the steepening trend from
outer-end to inner-end, indicating backflow of plasma in the lobes.  The
extended X-ray emission characterized by an absorbed power-law with photon index
$\sim$ 1.86 favours inverse-Compton scattering of the Cosmic Microwave
Background (ICCMB) photons as the plausible origin.  Using both X-ray and radio
fluxes under the assumption of ICCMB we estimate the magnetic field in the
lobes to be 3.3 $\mu$G.  The magnetic field estimate based on energy
equipartition is $\sim$ 3.5 $\mu$G.  Our work presents a case study of a rare
example of a GRG caught in dying phase in the distant Universe. 

\end{abstract}




\section{Introduction}

The radio morphology of a radio galaxy typically consists of a core, a
pair of highly collimated jets and lobes formed by the supply of relativistic
plasma from Active Galactic Nuclei \citep[AGN;][]{scheuer74,begelman84,bridle84}.
A less common population of radio galaxies with total linear size larger than
$\sim$ 0.7 Mpc is referred as Giant Radio Galaxies
\citep[GRGs; e.g.][]{saripalli05}.  After an active phase, which lasts for
$\sim$10$^{7}-10^{8}$ yr, the AGN activity stops or falls to very low
level such that the outflowing jets are no longer sustained. As a result, the
core, jets and hotspots on the lobes disappear.  However, the
lobes can be seen for relatively shorter period of time (10$^{6}-10^{7}$
yr) before they disappear due to radiative losses \citep{murgia11}.  Thus,
a relic radio galaxy resulting from the cessation of AGN activity represents a
short-lived final phase of radio galaxy evolution.  The short-lived phase makes
them a rare class of objects.  

The switch off of fresh particle injection leads to an exponential steepening
of the radio spectrum in the radio lobes, resulting in a much steeper
spectrum ($\alpha>1.3$) than that of normal radio galaxies
\citep{komissarov94}. The low surface brightness and very steep radio spectrum
makes relic GRGs ideally suited to be detected and studied at low frequencies.
However, despite concerted efforts to search for relic radio galaxies,
hitherto, only a handful of such objects have been reported in the
literature \citep[e.g.][]{cordey87, venturi98, murgia05, murgia11,
hurley-walker15}. In this paper, we report the discovery of a relic GRG 
J021659-044920 using deep 0.325 and 1.4 GHz observations carried out at the
Giant Metrewave Radio Telescope (GMRT) and the Very Large Array (VLA),
respectively. 

The reported relic GRG lies in the {\it XMM}-Large Scale Structure (XMM-LSS) field.
The existing X-ray data from {\it XMM$-$Newton} show extended X-ray emission nearly
co-spatial with the radio lobes. A variety of physical processes can give rise
to extended X-ray emission such as thermal emission from shocks, synchrotron
radiation and inverse-Compton (IC) scattering of the ubiquitous Cosmic
Microwave Background (CMB) photons. Indeed, IC scattered CMB
photons have been detected in the form of diffuse, extended X-ray emission in several
radio galaxies at redshift $z > 1$ [e.g. 4C41.17 \citep{scharf03}, 3C 294
\citep{fabian03}, 6C 0905+39 \citep{blundell06, erlund08}, 4C 23.56
\citep{johnson07}, and HDF-130 \citep{fabian09}, 3C 469.1 and MRC 2216--206
\citep{laskar10}].  However, given the dearth of relic GRGs and their X-ray
observations, X-ray properties of relic GRGs are poorly studied. For example,
only 2 out of 11 relic radio galaxies in \citet{parma07} and three out of five
relic radio galaxies in \citet{murgia11} sample have X-ray studies. For the relic GRG
J021659-044920, we use radio and X-ray observations to study its properties
e.g., morphology, dominant energy loss processes, energetics and magnetic field
strength.

In Section~\ref{obs}, we discuss the multiwavelength data used in our study.  In
Section~\ref{Radioproperties}, we present the radio morphology and spectral
properties of the source. The X-ray emission properties are discussed in
Section~\ref{xrayproperties}. In Section~\ref{radio-xray-connection}, we
discuss the connection between the radio and X-ray emission. We estimate
the magnetic field strength and the spectral age of the relic lobes in
Section~\ref{magfield-age}. We summarize our results in Section~\ref{summary}. 

Throughout the paper, spectral index is defined as $S_{\nu} \propto
\nu^{-\alpha}$. We use the {\it Wilkinson Microwave Anisotropy Probe}9 cosmology with $H_{0} = 69.32$ $ \textrm{km\
s}^{-1}\textrm{Mpc}^{-1}$, $\Omega_{\rm m} = 0.29$ and $\Omega_{\Lambda} = 0.71$.
All the frequencies are in the observed frame unless mentioned otherwise.

\section{Observations and data analysis}
\label{obs}

The relic GRG J021659-044920 centred at RA $02^{\rm{h}}~16^{\rm{m}}~59^{\rm{s}}$
and Dec. $-04^{\circ}~49^{\prime}~20^{{\prime}{\prime}}.6$ (J2000) is located in
a smaller subfield of the {\it XMM}-LSS field, known as the Subaru X-ray Deep Field
\citep[SXDF;][]{ueda08}.  Both SXDF and {\it XMM}-LSS have excellent multiwavelength
data coverage in X-ray, optical, near-infra red (IR), mid-IR and radio bands
\citep{simpson06,vardoulaki08,mauduit12}.  The details of the data used in our
study are given below.

\subsection{Radio data}
\label{radioobs}
 
\subsubsection{0.325 GHz data}

We use a GMRT observation at 0.325 GHz to study the low-frequency radio
properties of the relic GRG J021659-044920.  We have carried out deep 0.325 GHz
GMRT observations of the {\it XMM}-LSS field covering $\sim$ 12 deg$^{2}$ over
16-pointings with 32 MHz bandwidth. The GMRT observations are taken in
semisnapshot mode to optimize the $uv$-coverage.  The GMRT map has average rms
noise of $\sim$ 150 $\mu$Jy beam$^{-1}$ and angular resolution
9.35 arcsec $\times$ 7.38 arcsec. In
Figure~\ref{x-rayimages} (left-hand panel), we present the grey-scale image of
the relic GRG studied in this paper.  The details of the 0.325 GHz GMRT
observations will be presented in Wadadekar et al. (in preparation).  The
0.325 GHz GMRT observations in combination with the existing multiwavelength
data in the {\it XMM}-LSS field, have already been used to study the population of
ultrasteep spectrum radio sources \citep{singh14} and the radio-FIR correlation for
distant star-forming galaxies \citep{basu15}.

\subsubsection{0.240 GHz and 0.610 GHz data}

The {\it XMM}-LSS field was observed at 0.24 and 0.61 GHz using the GMRT by
\citet{tasse07}. The 0.240 and 0.610 GHz observations cover 18.0 and 12.7 deg$^{2}$
with rms noise $\sim$2.5 and $\sim$0.3 mJy beam$^{-1}$ and angular resolutions of
14.7 arcsec and 6.5 arcsec, respectively.
The Integrated flux densities of the relic GRG reported in \citet{tasse07} are
169.1$\pm$20.2 and 20.2$\pm$2.7 mJy at 0.240 and 0.610 GHz, respectively.
In a deeper GMRT observation of the SXDF at 0.610 GHz covering 0.5 deg$^{2}$
with map rms noise of $\sim$ 60 $\mu$Jy beam$^{-1}$, at a resolution of
6.8 arcsec $\times$ 5.4 arcsec by
\citet[][]{vardoulaki08}, the total flux density of the source is found to be
43.3 mJy. This clearly shows that the low surface brightness emission is missed
in the shallow observations reported in \citet{tasse07}.  We therefore use
the 0.610 GHz flux density given in \citet{vardoulaki08} for our analysis.

\subsubsection{1.4 GHz data}

\begin{figure*} \begin{centering} \includegraphics[scale=0.4]{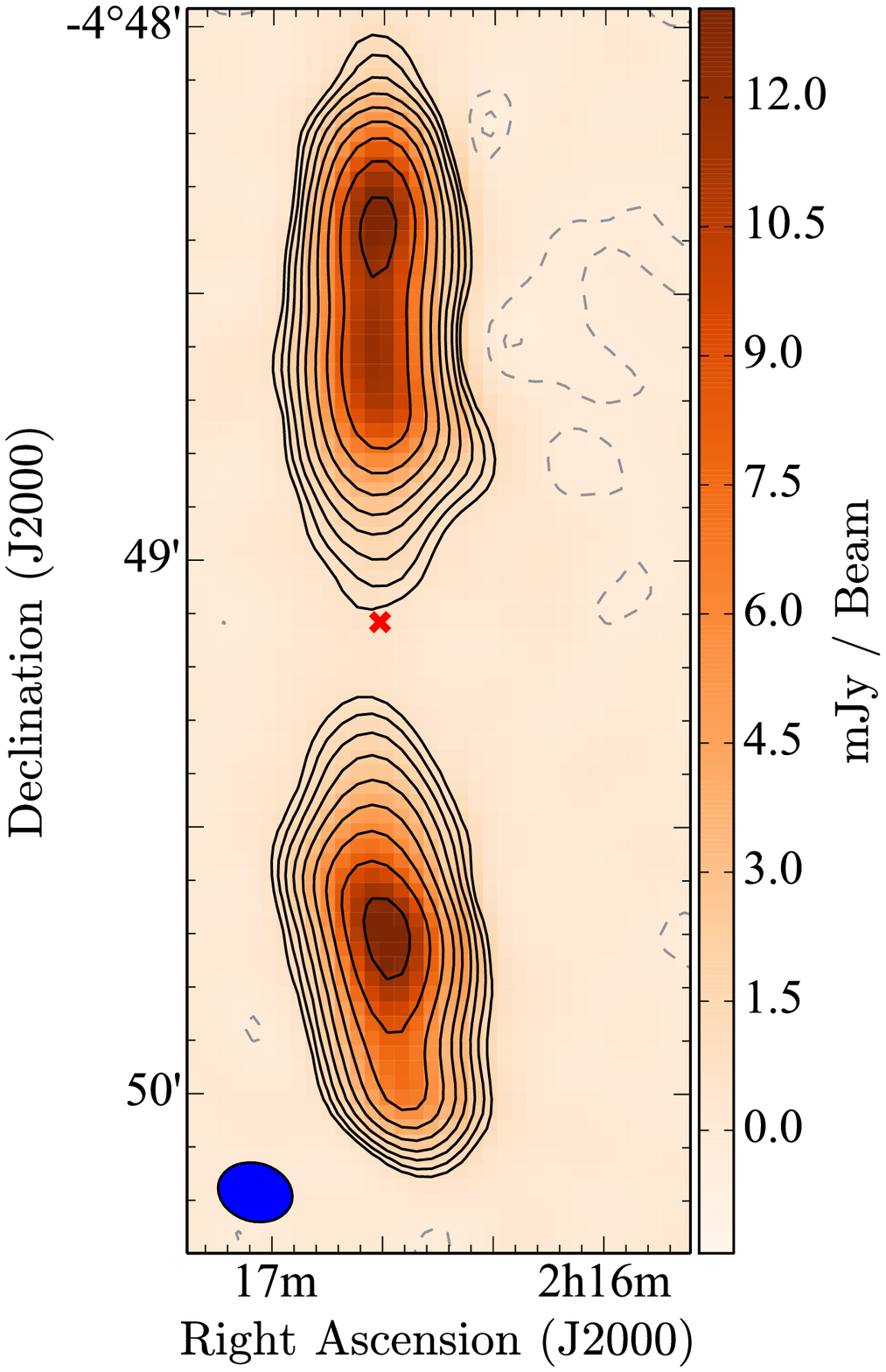}
\hspace{5pt}
\includegraphics[scale=0.4]{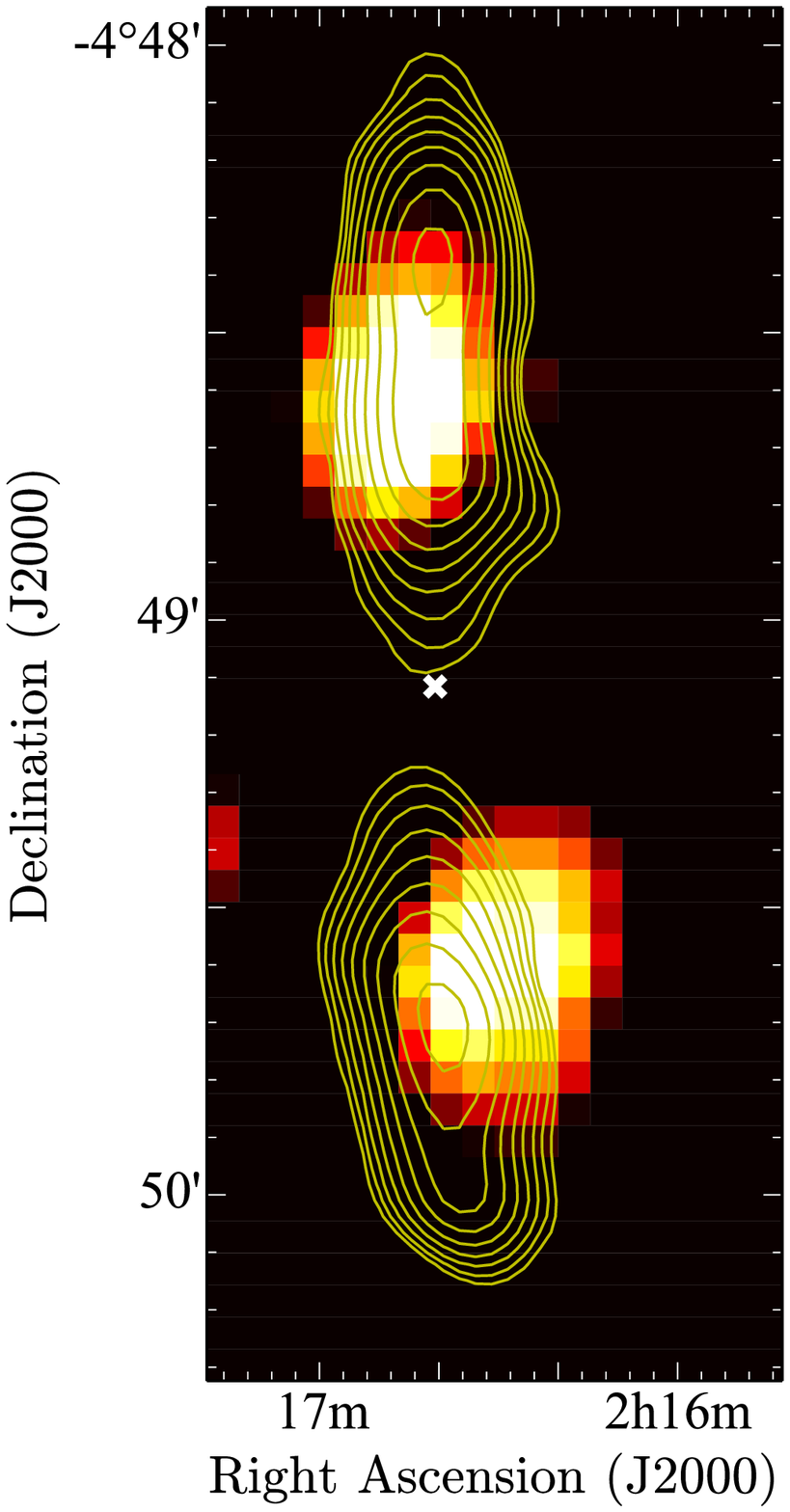}
\hspace{5pt}
\includegraphics[scale=0.4]{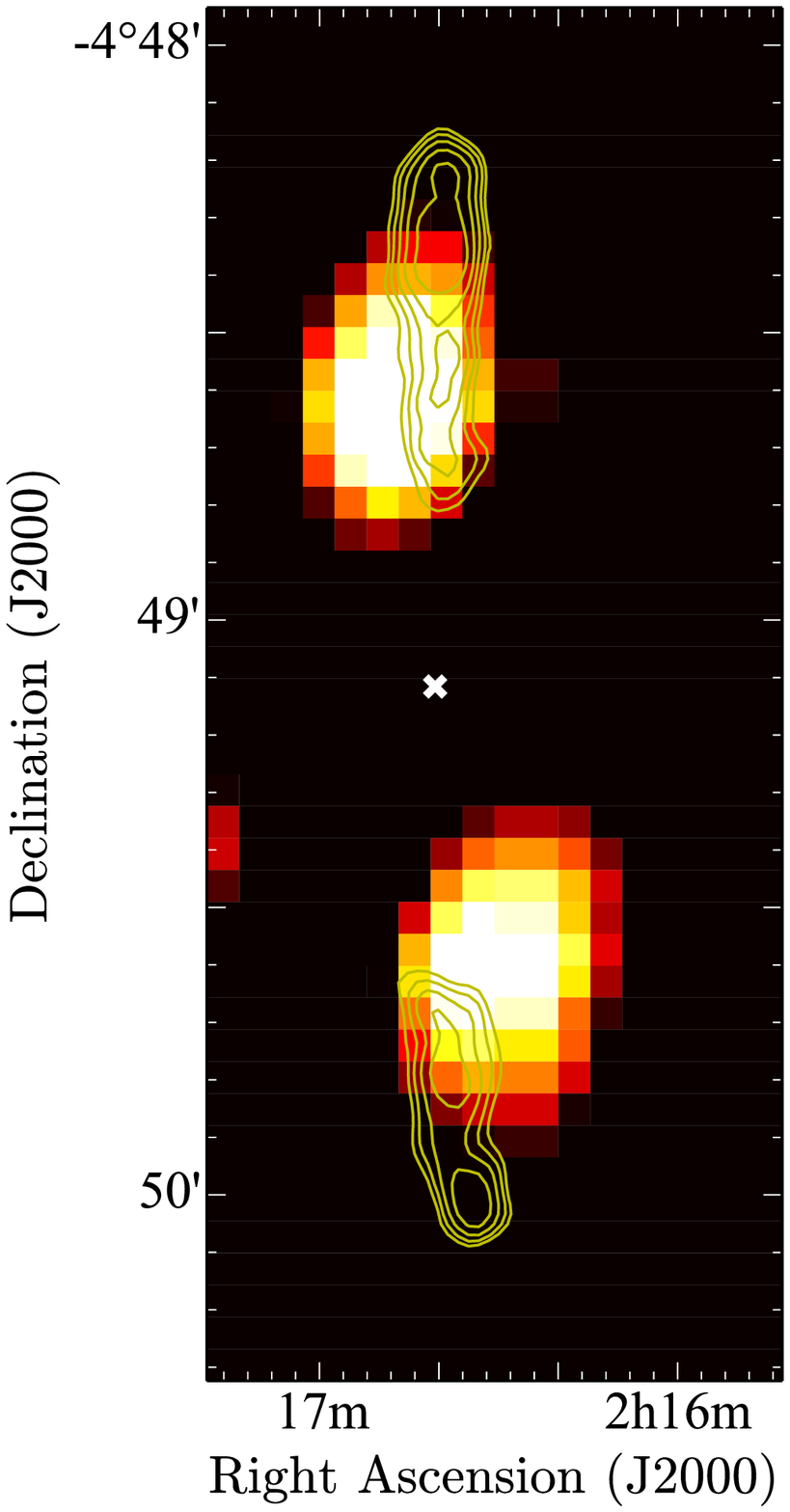}

 \caption{Left-hand panel: grey-scale image of the relic GRG
J021659-044920 at 0.325 GHz observed using the GMRT having angular resolution of 
9.35 arcsec $\times$ 7.38 arcsec. The map rms noise
is $\sim$ 150 $\mu$Jy beam$^{-1}$. Overlaid are the contours
at 0.325 GHz starting from 5$\sigma$ increasing in steps of $\sqrt2\times\sigma$.
The dashed grey contours are at $-3\sigma$ and $-2\sigma$. 
Middle Panel: grey-scale image of the soft X-ray band (0.3--2.0 keV) EPIC PN
from the {\it XMM$-$Newton}. Overlaid are the 0.325 GHz contours as in the left-hand panel. Right-hand panel:
the soft X-ray band image overlaid with the 1.4 GHz contours starting from
5$\sigma$ increasing in steps of $\sqrt2\times\sigma$. The 1.4 GHz observation
has angular resolution of $\sim$ 5.0 arcsec $\times$
4.0 arcsec and rms noise of $\sim$ 20 $\mu$Jy beam$^{-1}$. In both middle and right-hand panel, the X-ray image is
smoothed with a Gaussian of kernel radius 5.0 arcsec. The marker
`x' indicates the position of the host galaxy.}
 \label{x-rayimages}
\end{centering}
\end{figure*}

A deep 1.4 GHz image of the source, observed using the VLA, is available from
\citet{simpson06} who has surveyed the SXDF field covering 0.8 deg$^{2}$. The
VLA observations were made using B- and C-array configuration and render rms
noise of $\sim20~\mu$Jy beam$^{-1}$, at $\sim$ 5.0 arcsec $\times$
4.0 arcsec angular resolution. We show the 1.4 GHz contours in
Figure~\ref{x-rayimages} (right-hand panel).

\subsubsection{Missing flux density of radio emission} \label{missingflux}

The total flux density from a large angular size source can be underestimated
due to under sampling of the $uv$-plane at the shortest $uv$-	distance.  This
could be significant at higher radio frequencies. An interferometer with the
shortest baseline $D_{\rm min}$ can detect all the flux from angular scales
less than $\sim 0.6\lambda$/$D_{\rm min}$, provided the \textit{uv}-plane is
densely sampled at the shortest spacings. Here, $\lambda$ is the observing
wavelength. The total angular extent of the GRG is $\sim$2.4 arcmin at 0.325
GHz, and can be well sampled by baselines having $uv$-length $\gtrsim1$
k$\lambda$. This is not an issue with both the GMRT and VLA
observations at 0.325 and 1.4 GHz, as
the shortest baseline starts from $\sim$ 0.15 and
$\sim0.17~\rm{k} \lambda$, respectively. Hence, we do not expect
any missing flux density at 0.325 and 1.4 GHz and all the structures are well
recovered. Additionally, we compared the flux density at 1.4 GHz with
NRAO VLA Sky Survey (NVSS). They were found to be comparable indicating
negligible missing flux density.

Further, we note that the 0.610 GHz flux density measured by \citet{vardoulaki08}
used baselines above 1.5 k$\lambda$ for their analysis. Thus, the flux density of
43.3 mJy at 0.61 GHz could be underestimated due to missing flux and hence is
lower than the actual flux density.

\subsection{X-ray data}

We used {\it XMM$-$Newton} archival data with observation ID 0112372001. The
data were taken in prime-full window mode using thin filter on 2003 January 7,
and consist of 25.63 ks EPIC PN and 27.35 ks of EPIC MOS data.  EPIC
comprises a set of three X-ray CCD cameras i.e., two MOS cameras (Metal Oxide
Semi-conductor CCD arrays) and one PN camera (pn CCD array). The EPIC cameras
perform sensitive imaging observations in the energy range 0.15---15 keV
with spatial resolution 4.0 arcsec --- 6.0 arcsec
and spectral resolution (E/$\Delta$E) $\sim$ 20 --- 50 over the telescopes field
of view of 30 arcmin \citep{struder01,turner01}.
The sensitivity of EPIC MOS/PN sharply declines above 10 keV
resulting in inadequate counts above 10 keV for these observations.
Figure~\ref{x-rayimages} shows spatially smoothed EPIC-PN soft band (0.3 --- 2.0
keV) X-ray image of the relic GRG.  Overlaid are 0.325 GHz (middle panel) and
1.4 GHz (right-hand panel) radio contours. The soft X-ray image is smoothed with a
Gaussian of kernel radius 5.0 arcsec. X-ray
emission from extended diffuse double lobes of the relic GRG are clearly seen
and are nearly co-spatial with 0.325 GHz lobe emission.  We describe the
details of X-ray data analysis in Section~\ref{xrayproperties}.

\begin{table*}
\begin{centering}
\caption{AB magnitudes of the host galaxy in the optical and IR bands.}
\begin{tabular}{l|llllllllllll}
 \hline
 Band & {\it B} & {\it V} & \it R & \it i${'}$ & \it z${'}$ & \it J & \it H & \it K & 3.6$\mu$m & 4.5$\mu$m & 5.8$\mu$m & 8$\mu$m \\
 \hline
 Magnitude & 25.08 & 24.62 & 23.96 & 22.98 & 22.02 & 20.30 & 19.23 & 18.13 & 17.83 & 17.91 & 18.66 & 19.23\\
 \hline
\label{opt-ir-fluxes}
\end{tabular}
\end{centering}
\medskip
\\ {\begin{flushleft} \textit{Note}. Optical band magnitudes ({\it B, V, R, i${'}$, z${'}$}) are from 
Subaru data~\citep{simpson06}, near-IR fluxes ({\it J, H, K}) are from the UKIDSS-UDS catalogue and
mid-infrared fluxes (3.6, 4.5, 5.6, 8$\mu$m) were measured from the
SpUDS images within a fixed circular aperture of 6.0 arcsec diameter.
\end{flushleft}}
\end{table*}

\subsection{Optical, near-IR data and host galaxy identification}

\begin{figure}
\begin{centering}
 \includegraphics[scale=0.4]{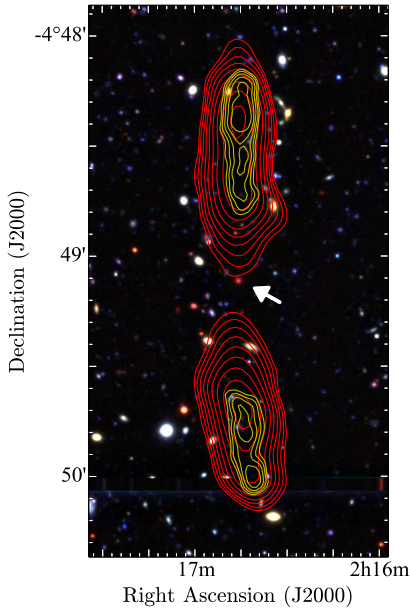}
 \caption{ 0.325 GHz (in red) and 1.4 GHz (in yellow) radio contours overlaid
on the three-colour optical RGB image.  RGB three-colour image obtained by combining
Subaru {\it B} band (blue), {\it R} band (green) and {\it z${'}$} band (red) images. The extreme red
colour of the host galaxy is apparent in RGB image. 
The arrow points to the host galaxy. Both 0.325 and 1.4 GHz contours
start from 5$\sigma$ and increase in steps of $\sqrt{2}~\times~\sigma$.} 
 \label{irimages}
\end{centering}
\end{figure}

\citet{simpson06} present optical identifications for all radio sources detected
in the 1.4 GHz imaging of the SXDF.  The optical identifications of
radio sources is based on the ultra-deep {\it B, R, i${'}$, z${'}$}
Suprime-Cam/Subaru images.  The optical identifications were made by overlaying
the radio contours on to a true-colour {\it B, R, z${'}$} image. The relatively high
resolution of the 1.4 GHz radio map allowed unambiguous identification with
optical sources as there was typically only one source within an arcsecond of
the peak of the radio emission, even at extremely faint optical magnitudes. For
sources composed of multiple radio components without core emission, such as
our case, an identification was sought at the centre of the source. The relic
GRG without a radio core presented in this study has been assigned a reliable
optical counterpart at optical position RA = 02$^{\rm h}$ 16$^{\rm m}$ 59$^{\rm s}$.064, Dec. = $-04^\circ$
49$^{\prime}$ 20$^{{\prime}{\prime}}$.85 with spectroscopically measured
redshift $z_{\rm spec} \sim$ 1.325 \citep[see table~3 in][]{simpson06}. The identified
host galaxy of the relic GRG is a very red galaxy ({\it R $-$ z${'}$} = 2.0) and
resides almost exactly midway between the radio lobes.  Figure~\ref{irimages}
shows overlay of 0.325 GHz radio contours on the optical three-colour RGB image
obtained by combining Subaru {\it B} band (blue), {\it R} band (green) and {\it z${'}$} band
(red) images.  Optical magnitudes of the host galaxy in {\it B, V, R, i${'}$} and
{\it z${'}$} bands from \citet{simpson06} are listed in Table~\ref{opt-ir-fluxes}.
The host galaxy is also detected in near-IR ({\it J}, {\it H} and {\it K} bands) and mid-IR (3.6,
4.5, 5.8 and 8 $\mu$m) bands observed as a part of UK Infrared Deep Sky Survey
(UKIDSS) Ultra Deep Survey \citep[UDS;][]{lawrence07} and {\it Spitzer} UKIDSS Ultra
Deep Survey (SpUDS; PI: James S. Dunlop)\footnote{\url{http://irsa.ipac.caltech.edu/data/SPITZER/SpUDS/}}, respectively.

\subsection{Photometric redshift of the host galaxy}

The spectroscopic redshift of the extremely red host galaxy provided by
\citet{simpson06} is based on a low S/N spectrum from FOCAS/Subaru.  In order
to check the reliability of spectroscopic redshift, we estimated photometric
redshift using 12 band photometric data ({\it B, V, R, i${'}$} and {\it z${'}$} band from
Subaru, {\it J, H, K} band from UKIDSS and 3.6, 4.5, 5.0 and 8.0 $\mu$m from SpUDS).
We estimated the photometric redshift ($z_{\rm phot}$) of the host by fitting
template SEDs to the photometric data using the publicly available photometric
redshift estimation code {\sc eazy}\footnote{\url{http://www.astro.yale.edu/eazy/}}
\citep{brammer08}.  We obtain $z_{\rm phot} = 1.26^{+0.10}_{-0.03}$, similar to
the spectroscopic redshift, $z_{\rm spec}$ = 1.325, within $1\sigma$.  Our
$z_{\rm phot}$ is also consistent with $z_{\rm phot}$ in the range 0.98 ---
1.174 reported by \citet{vardoulaki08} using relatively shallow mid-IR data
from SWIRE. Our estimated $z_{\rm phot}$ independently confirms the $z_{\rm
spec}$ estimate to be reliable. We use $z_{\rm spec}=1.325$ for our
calculations.

\section{Radio properties}
\label{Radioproperties}

\subsection{Radio morphology}

\begin{table*}
\caption{Radio flux densities and spectral indices of relic GRG J021659-044920.
Column 1: flux density at 1.4 GHz measured using \citet{simpson06} image.
Column 2 is 0.610 GHz flux density of the GRG taken from \citet{vardoulaki08}.
Column 3 is 0.325 GHz flux density measured using our 0.325 GHz image.
Column 4 is 0.240 GHz flux density of the GRG taken from \citet{tasse07}.
Columns 5, 6, 7 and 8 are spectral indices between 1.4 to 0.610 GHz, 1.4 to 0.325 GHz,
0.610 to 0.325 GHz and 0.325 to 0.240 GHz, respectively.
All flux densities are integrated within 5$\sigma$ contours.}
\begin{tabular}{ccccccccc}
 \hline
  & 1.4 GHz & 0.610 GHz & 0.325 GHz & 0.240 GHz & ${\alpha}_{\rm 0.610~GHz}^{\rm 1.4~GHz}$ & ${\alpha}_{\rm 0.325~GHz}^{\rm 1.4~GHz}$ 
& ${\alpha}_{\rm 0.325~GHz}^{\rm 0.610~GHz}$ & ${\alpha}_{\rm 0.240~GHz}^{\rm 0.325~GHz}$ \\

  & (mJy) & (mJy) & (mJy) &  (mJy) &   &  &  &  \\
  & (1) & (2) & (3) & (4) & (5) & (6) & (7) & (8) \\
 \hline
 North lobe & 5.80$\pm$0.10 & --- & 79.8$\pm$5.20 &   &  & 1.79$\pm$0.11 &  &  \\
 South lobe & 3.80$\pm$0.09 & --- & 60.2$\pm$4.52 &   &  & 1.89$\pm$0.12  &  &   \\
 Total & 9.60$\pm$0.14 & $>$43.3 & 140$\pm$6.9 & 169.1$\pm$20.2 & $>1.81$ & 1.83$\pm$0.08 & $<1.86$ & 0.6$\pm$0.5  \\
 \hline
\label{radio_fluxes}
\end{tabular}
\end{table*}

We detect the two radio lobes in both the 0.325 and 1.4 GHz images, while the
AGN core, jets and/or hotspots remain undetected.  The non-detection of AGN
core, jets and hot-spots classifies this source to be a relic radio galaxy
\citep[e.g.,][]{murgia11, saripalli12}.

At 0.325 GHz, the total integrated flux density of the relic GRG J021659-044920
is 140$\pm$6.9 mJy with northern and southern lobes having integrated flux
densities of 79.8$\pm$5.2 mJy and 60.2$\pm$4.5 mJy, respectively.  In Table~2,
we list the flux densities of the GRG at various radio frequencies.  The total
angular extent at 0.325 GHz (within 5$\sigma$ contour) is 2.4 arcmin and
corresponds to a projected linear size of $\sim$1.2 Mpc at redshift 1.325.
The projected linear sizes of northern and southern lobes at 0.325 GHz are
$\sim$ 600 kpc (1.18 arcmin) and $\sim$ 510 kpc (1 arcmin),
respectively.  At 1.4 GHz the source size is relatively smaller with total
end-to-end projected linear size of 1.1 Mpc (2.26 arcmin).  The total
integrated flux density at 1.4 GHz is $9.60\pm0.14$ mJy with northern and
southern lobes having $5.8\pm0.1$ mJy and $3.80\pm0.09$ mJy, respectively.
Using 5$\sigma$ rms noise as the upper limit for the undetected AGN core, we
estimate the core-to-lobe flux density ratio to be $\lesssim 1$ and $\lesssim
5$ per cent at 1.4 and 0.325 GHz, respectively. This low value of
core-to-lobe flux density ratio is similar to other relic GRGs \citep[see
e.g.,][]{hurley-walker15}.

\subsection{Radio spectrum and spectral index map}
\label{spec-index-map}

\begin{figure*} 
\begin{centering}
\begin{tabular}{cc}
{\mbox{\includegraphics[width=8cm]{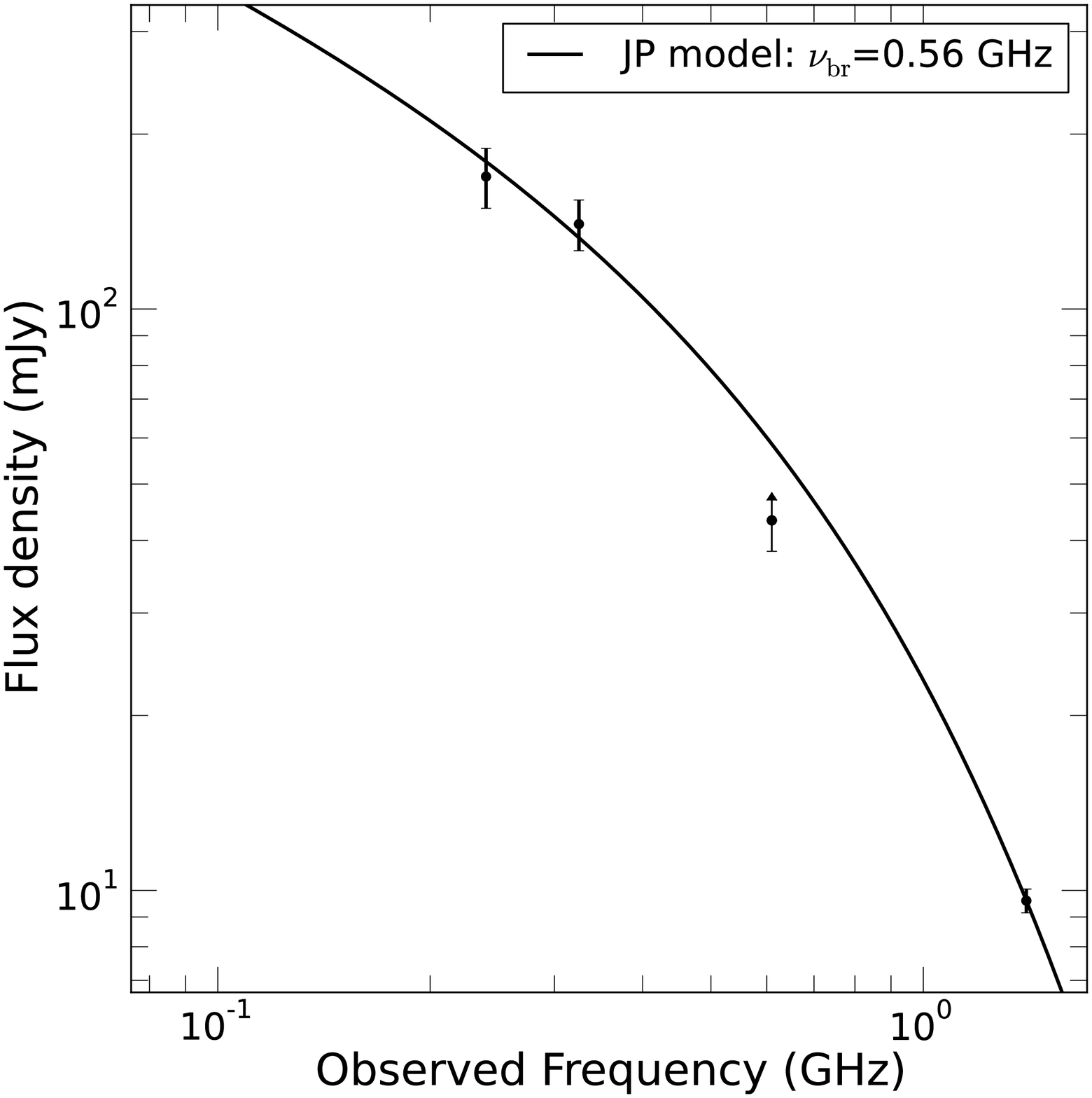}}} &
{\mbox{\includegraphics[width=8cm]{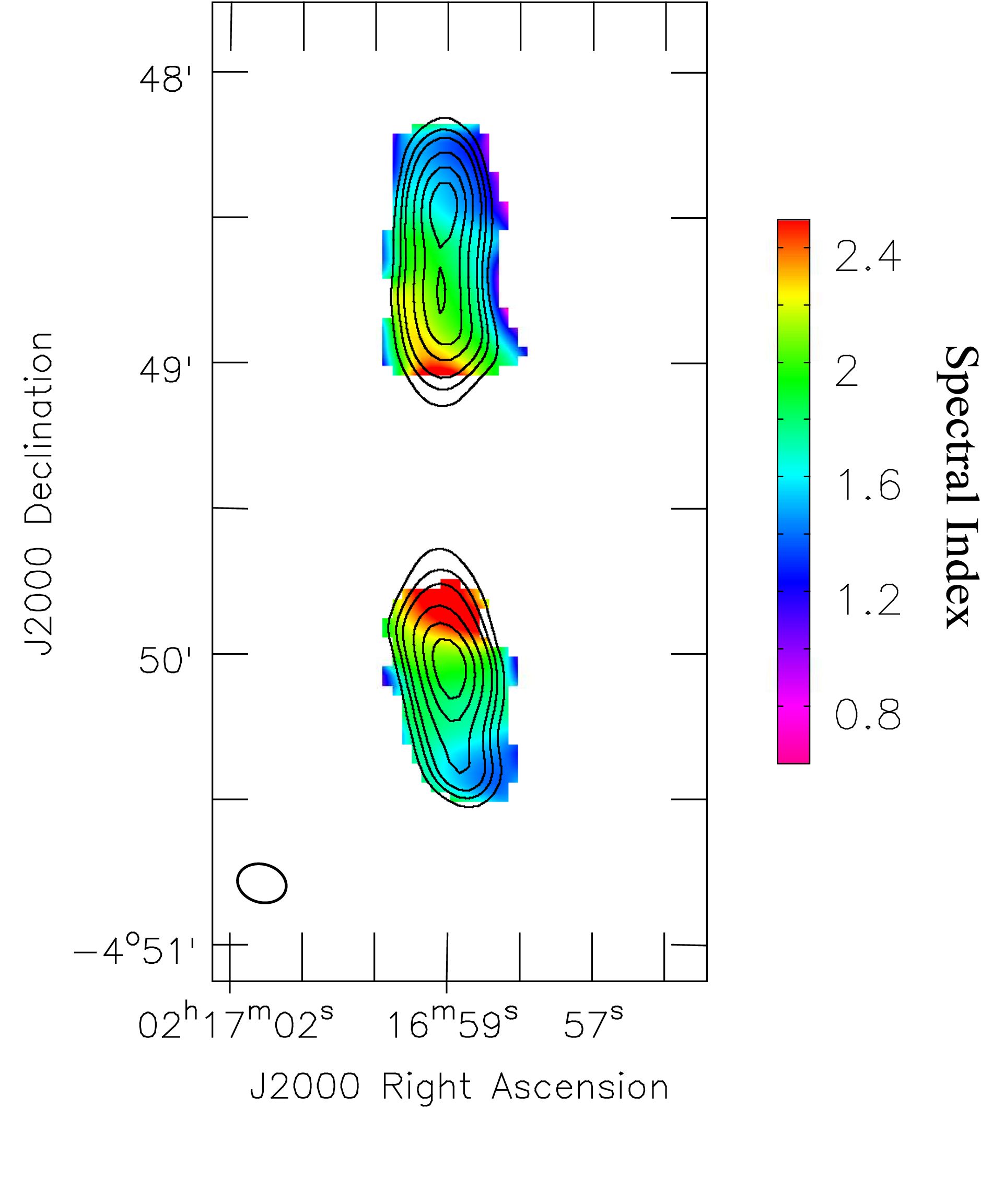}}}\\
\end{tabular}
\end{centering}
\caption{Left-hand panel: integrated radio continuum spectrum
between 0.24 and 1.4 GHz of the relic GRG.  The spectrum is fitted using the JP model
for synchrotron radiative loss. Right-hand panel: spectral index map
of the GRG estimated between 0.325 and 1.4 GHz using 3$\sigma$ flux density levels.
Overlaid are the 0.325 GHz total intensity contours.}
\label{spind}
\end{figure*}

In Figure~\ref{spind} (left-hand panel), we show the radio spectrum for the GRG
J021659-044920 between 0.24 and 1.4 GHz.  We list the radio spectral indices
computed across 0.240, 0.325, 0.610 and 1.4 GHz in Table~\ref{radio_fluxes}.  
The solid line shows the best-fitting spectrum fitted by a power-law with
exponential cutoff. We describe the spectral modelling in Section~3.3. It is
evident that the radio spectrum is convex and steepens sharply above 0.325 GHz.
The source shows strong spectral evolution from 0.240 to 1.4~GHz. The radio
spectral index estimated between 0.325 and 1.4 GHz is $\alpha$ $\sim$ 1.8,
consistent with fading lobes without injection of fresh particles from AGN. In
the literature, radio spectral curvature is computed as the difference between
high-frequency spectral index and low-frequency spectral index (i.e,
${\alpha}_{\rm high} - {\alpha}_{\rm low}$). ${\alpha}_{\rm high} -
{\alpha}_{\rm low}\geq 1.0$ has been used as a criterion to selected relic
(dying) radio galaxies \citep[e.g.,][]{murgia11}.  With this criterion our
source having radio spectral curvature (${\alpha}_{\rm 0.610~GHz}^{\rm 1.4~GHz}
- {\alpha}_{\rm 0.240~GHz}^{\rm 0.325~GHz}$) $> 1.19$, can be categorized as a
dying radio galaxy.  In the active phase, radio galaxies generally show radio
spectra represented by a power-law over a wide range of frequencies.  
However, in the absence of fresh particle injection during the relic phase, the
energy spectrum of the synchrotron emitting relativistic electrons develop an
exponential cut-off towards higher energies due to synchrotron and/or IC losses.
This leads to an cut-off in the high frequency radio
spectrum. Therefore, the fading lobes in relic GRG are expected to show very
steep ($\alpha$ $>$ 1.3) and convex radio spectra, characteristic of electron
population that has radiated away much of the energy \citep{komissarov94}.

In Figure~\ref{spind} (right-hand panel), we present the spectral index map computed
between 0.325 and 1.4 GHz using pixels above 3$\sigma$ at both the frequencies.
The 0.325 GHz emission extends
well beyond 1.4 GHz emission in the lobes.  Spectral index varies from 1.4
to 2.5 with gradual steepening in the lobes from outer extremities to inner regions
towards the host galaxy. The average spectral index of the northern lobe is
$\sim1.8$, and that of the southern lobe is $\sim1.9$ (see Table~\ref{radio_fluxes}).
The spectral index is extremely steep relative to typical low-$z$ radio galaxies
where it is $\sim 0.7$ \citep{blundell99}.  However, the variation of spectral
index with distance from the nucleus in the lobes is consistent with the
backflow model of the lobes \citep{leahy84,leahy89}. According to this model,
pressure in the hotspot re-accelerates the post-shock material back towards the
core, creating a backflow. Thus, the part of the lobe closer to the host galaxy
consists of an older population of cosmic ray electrons (CREs) that are backflowed
and makes the radio continuum spectrum steep in those regions.

\subsection{Energy losses}

The relativistic electrons in the radio lobes undergo energy losses that
modifies the power-law radio spectrum.  The dominant energy loss processes
are synchrotron, IC and adiabatic cooling. Synchrotron and
IC losses have the effect of steepening the spectrum and leads to a cutoff at
higher radio frequencies ($\gtrsim 1$ GHz), while adiabatic losses affect the
normalization of the spectrum but not its curvature. In the following we
suppose that the radiative losses are the dominant energy loss process for
the relativistic electrons. We neglect the effect of the expansion losses
as this is a dying source and perhaps in equilibrium with external pressure.

Both, synchrotron and IC losses affect the spectrum in similar ways, whereby,
the spectrum is smoothly steepened at higher frequencies.  The spectrum is
characterized by a break frequency, $\nu_{\rm br}$, below which the spectrum
remains a power-law with a spectral index identical to the injection spectral
index ($\alpha_{\rm inj}$).  Above $\nu_{\rm br}$, the form of the steepening
depends on the mechanism of particle injection. For steady continuous injection
of relativistic electrons (CI model), the spectrum steepens by 0.5
\citep{pacho70}. For single shot particle injection, the spectrum falls off as
a power-law with index $4\alpha_{\rm inj}/3 +1$ above the break frequency,
assuming constant pitch angle between electrons and magnetic field \citep[KP
model;][]{karda62, pacho70}. However, considering rapid isotropization of the
pitch angle distribution leads to an exponential cut-off above the break
frequency \citep[JP model;][]{jaffe73}.

\begin{figure} 
\begin{centering}
{\mbox{\includegraphics[width=9cm]{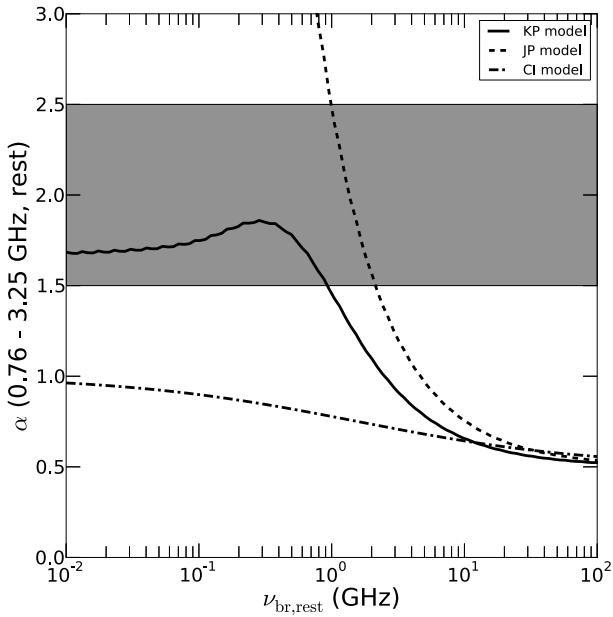}}} 
\end{centering}
\caption{Expected spectral index between rest-frame 0.76 and 3.25 GHz as a
function of the break frequency at rest frame, $\nu_{\rm br, rest}$ for the
various models of particle injection with $\alpha_{\rm inj}=0.5$. The shaded
region shows the observed range of spectral index in the lobes.}
\label{loss}
\end{figure}

In Figure~\ref{loss} we show the expected spectral index between rest-frame
0.76 and 3.25 GHz (corresponding to 0.325 and 1.4 GHz in the
observed frame) as a function of the break frequency in the rest frame
($\nu_{\rm br, rest}$) for the different models described above. We assumed an
injection spectral index $\alpha_{\rm inj}=0.5$, and produced synthetic radio
spectrum for varying $\nu_{\rm br, rest}$. The shaded region shows the observed
range of spectral index in the lobes. Clearly, the CI model cannot produce
the observed steepness and would require $\alpha_{\rm inj} > 1$, which is
generally not the case. The fact that the extended radio emission is very
bright at low radio frequencies and extends well beyond high frequency radio
emission together with the observed spectral index supports the fact that the
lobes consists of an old population of electrons that were injected into the
intergalactic medium (IGM) from a now discontinued jet activity.

From Figure~\ref{loss}, we find that the observed steepening of the spectral
index can be produced by the JP model. As per the JP model, the range of
spectral index suggests, $\nu_{\rm br,rest}$ to be in the range 1$-$2.2 GHz in
the rest frame. We therefore model the observed radio spectrum using the JP
model. However, we note that, the JP spectrum with a single $\nu_{\rm br}$ 
is better represented for localized regions. The total radio emission is a sum
of several JP spectra having a range of $\nu_{\rm br}$. Depending on the time 
since first injection, i.e., the total age of the source
($t_{\rm s}$) and the time elapsed since injection switch-off ($t_{\rm off}$), 
the $\nu_{\rm br}$ lies in the range $\nu_{\rm br, low}$ and $\nu_{\rm br, high}$, 
respectively. Here, $t_{\rm s} = t_{\rm off} + t_{\rm on}$, where, $t_{\rm on}$
is the duration of the active phase. The two break frequencies are related as, 
$\nu_{\rm br, high}=\nu_{\rm br, low}(t_{\rm s}/t_{\rm off})^2$ 
(see ~\citet{murgia11}). In the scenario, where the dying phase is 
longer than that of the active phase, i.e. $t_{\rm off} >> t_{\rm on}$, 
the two break frequencies overlap and the global spectrum can be represented
by a JP model consisting of a single break frequency. In our case,
the spectrum is not sampled enough to 
determine the two break frequencies. We therefore model the spectrum by 
a single $\nu_{\rm br}$ and approximate the JP spectrum as 
$S_{\nu} = A \nu^{-\alpha_{\rm inj}} {\rm e}^{-\nu/\nu_{\rm br}}$. Here, 
$A$ is the normalization at 1 GHz and $\alpha_{\rm inj}$ is the 
injection spectral index. Under this assumption, the $\nu_{\rm br}$ 
is determined by $t_{\rm s}$. We assume,
a typical value of $\alpha_{\rm inj}$ to be 0.5. We fit the above form
for the JP model to the observed radio continuum spectra between 0.24 and
1.4 GHz. For our data, we find $\nu_{\rm br}=0.57\pm0.02$ GHz, i.e,
$\nu_{\rm br, rest} = 1.32$ GHz and is consistent with what is expected
from Figure~\ref{loss}. The best-fitting radio spectrum is shown as the solid
line in Figure~\ref{spind} (left-hand panel).

\section{X-ray properties} \label{xrayproperties}

\begin{figure}
 \includegraphics[scale=0.347]{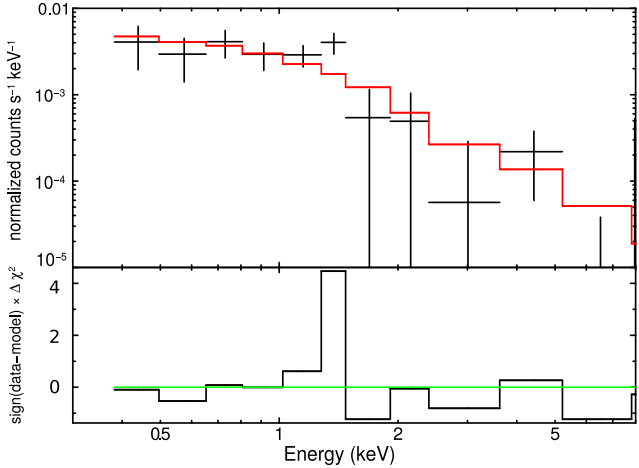}
 \caption{0.3 --- 10 keV {\it XMM$-$Newton} PN spectrum best fitted with an absorbed power-law, where absorbing column density is fixed to the Galactic value.
Solid red line and crosses (`+') represent fitted model and binned data points, respectively. Residuals are shown in the bottom panel.}
 \label{pnspectrum}
\end{figure}

\subsection{X-ray data reduction}

We performed X-ray data reduction using Science Analysis System ({\sc sas}) version
13.5 and followed standard procedure outlined in the {\it XMM$-$Newton} ABC Guide\footnote{\url{https://heasarc.gsfc.nasa.gov/docs/xmm/abc/}}.  We
processed raw Observations Data Files by using {\sc sas} tasks \texttt{emproc}
and \texttt{epproc}, for the MOS and PN data.  We use latest calibration files
to obtain calibrated and concatenated event lists.  These files were filtered
for flaring particle background time intervals using standard recommended
PATTERN, FLAG and energy filters to generate clean event lists.  From clean
calibrated data files we extracted source and background spectra.  Source
spectrum was extracted by choosing two elliptical regions covering X-ray
emission from each of the two lobes. The counts are not
sufficient to obtain robust spectral fits for individual lobes, and therefore,
we opted to extract source spectrum covering both the lobes. The background
spectrum is extracted using a source-free annular region in the same chip.  The
response matrix file and the ancillary response file were created using the {\sc sas}
tasks RMFGEN and ARFGEN, respectively. The spectra were grouped to
a minimum of 20 counts per bin in order to use ${\chi}^{2}$ fitting statistics.
The MOS and PN data are consistent with each other but PN data are slightly better
at higher energies $>$ 2.0 keV and therefore, we present spectral fitting using
the PN data only.

\subsection{X-ray spectral fitting and analysis}

To study the nature of X-ray emission from lobes we fitted the 0.3 --- 10 keV
EPIC PN spectrum using physically motivated models.  We use the X-ray spectral
fitting package {\sc xspec} version 12.8.2 \citep{arnaud96} to fit the X-ray
spectrum. Since the relic GRG has no AGN core emission, we fit the 0.3 --- 10
keV spectrum with a single absorbed power-law of the form, $M(E) = K {\rm e}^{-N_{\rm
H}{\sigma}(E)} E^{-{\Gamma}}$. Here, $K$ is the normalization of the emission
at 1 keV, $N_{\rm H}$ is the absorbing column density between source and
observer, $\sigma(E)$ is the photoelectric absorption cross-section (ignoring
Thomson scattering) and $\Gamma$ is the photon index of the power-law.  We get
$N_{\rm H}$ = 2.42$^{+5.31}_{-2.47}$ $\times$ 10$^{21}$ cm$^{-2}$, $\Gamma$ =
2.99$^{+3.13}_{-1.38}$ and ${\chi}^{2}$/dof = 7.7/10 $\sim$ 0.77. Keeping all
the parameters free does not constrain $\Gamma$.  However, by fixing the
absorbing column density equal to the Galactic value ($N_{\rm H}$ = 2.52
$\times$ 10$^{20}$ cm$^{-2}$) statistically improves the fit with
${\chi}^{2}$/dof = 9.7/11 $\sim$ 0.88 and better constrains $\Gamma$ to be
1.86$^{+0.49}_{-0.41}$. We use this value of $\Gamma$ in our further analysis.
The 0.3 --- 10 keV EPIC PN spectrum fitted with absorbed power-law model for
fixed $N_{\rm H}$ is shown in Figure~\ref{pnspectrum} and the best-fitting
parameters are listed in Table~\ref{xrayfitparams}.

To investigate the possibility of thermal origin of X-ray emission from hot
diffuse gas present in the cluster medium, we also fitted the 0.3 --- 10 keV
spectrum with a thermal plasma model ({\sc mekal}).  The {\sc mekal} model
represents an emission spectrum from hot diffuse gas based on the model
calculations of \citet{mewe86}.  The {\sc mekal} model with multiplicative
Galactic absorption component does not render a statistically good fit if all
the parameters are kept free. We tried fixing the metal abundance of the hot
gas in the cluster medium in the {\sc mekal} model from 0.1 to 1.5 of Solar
values in steps of 0.1 but found unphysical or unconstrained plasma
temperatures (see Table~\ref{xrayfitparams}).  Therefore, it is unlikely that
X-ray emission from the lobes are of thermal origin. In fact, the extended
X-ray morphology nearly co-spatial with the radio lobes precludes it from being
due to hot cluster gas.  \citet{murgia12} reported that relic radio galaxies
residing in cluster environments display large scale diffuse X-ray emission
having spherical morphology. And, the X-ray emission is characterized by
thermal emission from hot gas present in the cluster.  We note that
\citet{finoguenov10} present a list of X-ray emitting clusters in the SXDF but
the X-ray emission in our source is reported to be mainly from radio lobes.
Therefore, we conclude that the observed X-ray emission is non-thermal in
origin and perhaps originates due to IC scattering of the CMB photons (ICCMB).

\section{Connection between radio and X-ray emission}
\label{radio-xray-connection}

The co-spatial nature of the radio synchrotron and the soft X-rays emission
from the lobes suggests a fundamental connection between the emitting particle
populations within the lobes.  A possible scenario is IC
scattering of CMB photons by CREs having Lorentz
factors, $\gamma_e \gtrsim 10^3$, wherein the CMB photons are upscattered to
X-ray wavebands. Depending on the magnetic field strengths, CREs with $\gamma_e
\sim 10^2$ --- $10^5$ give rise to synchrotron emission in the radio wavebands
between $\sim$ 10 MHz and 100 GHz.  However, due to radiative losses, the
lifetime of relativistic electrons are inversely proportional to $\gamma_e$.
Thus, the radiative lifetimes of (highly energetic) radio synchrotron-emitting
electrons are shorter than the (less energetic) electrons which give IC
emission in the X-ray band. As a consequence, IC emission always traces an
older population of particles which may be more diffuse and spatially
non-coincident with the radio emission. This gives rise to smaller extent
and larger off-set between the radio emission from the lobes at 1.4 GHz and the
X-ray emission (see Figure~\ref{x-rayimages}). This is because, at 1.4
GHz the emission is produced by a population of relatively short-lived higher
energy electrons as compared to the 0.325 GHz emission.

\begin{table}
\begin{centering}
\caption{Best-fitting parameters for the power-law and {\sc mekal} models. 
Errors represent 90 per cent confidence interval on all parameters. }
\begin{tabular}{llc}
 \hline
 Model & Parameter & Value \\
 \hline
 power-law & K$_{\rm norm}$ & 5.29$^{+1.30}_{-1.30}$ $\times$ 10$^{-6}$ \\ 
           & {\it N}$_{\rm H}$ (cm$^{-2}$) (fixed)    & 2.52 $\times$ 10$^{20}$ \\
           & $\Gamma$ & 1.86$^{+0.49}_{-0.41}$ \\
  & $\chi^{2}$/dof & 9.7/11 \\
  & 0.3 --- 10 keV flux $^{1}$ & 3.25 $\times$ 10$^{-14}$ \\
  & 2.0 --- 10 keV flux $^{2}$ & 1.50 $\times$ 10$^{-14}$ \\
  &  & \\
 {\sc mekal} &     &    \\
             &  K$_{\rm norm}$  & 7.90$^{+1.09}_{-1.93}$ $\times$ 10$^{-5}$ \\
             & $kT$ (keV) & 4.2$^{+15.3}_{-2.3}$ \\
  & nH (cm$^{-3}$) & 1.17 $\times$ 10$^{-2}$ \\
  & Abundance & 0.6 \\
  & $\chi^{2}$/dof & 9.27/10 \\
 \hline
\label{xrayfitparams}
\end{tabular}
\end{centering}
\\
$^{1}$ Observed frame unabsorbed flux in erg cm$^{-2}$ s$^{-1}$.\\
$^{2}$ Rest frame unabsorbed flux in erg cm$^{-2}$ s$^{-1}$.\\
\end{table}

The equivalent magnetic field for IC scattering with CMB,
$B_{\rm IC}$, is given by $\approx3.25(1+z)^2~\mu$G.  For the GRG, $B_{\rm IC}
\approx 17.5~\mu$G is significantly stronger than that of the estimated
magnetic field of $\sim3.3~\mu$G (see Section~\ref{Magneticfield}) indicating IC losses significantly dominates over the synchrotron losses. This is
because the IC cooling time-scale for electrons radiating at 1.4 GHz
(i.e., 3.25 GHz in the rest frame) in a magnetic field strength of $3.3~\mu$G
is $\sim 4.8\times 10^6$ yrs, significantly lower than that of the synchrotron
cooling timescale of $\sim1.2\times10^8$ yr. Thus, the radio emission from the
lobes of the GRG are strongly affected by IC losses.

The power-law nature of the X-ray emission from the lobes can arise due to
synchrotron emission from highly energetic electrons having $\gamma_e \sim
2.7\times10^8$ at 2 keV (4.65 keV in the rest frame). Such electrons have radiative
(synchrotron $+$ IC) lifetime of $\sim 2.5 \times 10^2$ yr in a
magnetic field of $\sim$ 3.3 $\mu$G at $z=1.325$. This is about four orders
of magnitude lower than the estimated spectral age of $\sim$ $8 \times 10^6$
yr in the lobes (see Section~\ref{relicage}).
We therefore conclude that, the X-ray emission is a direct consequence of IC
scattering of the CMB photons with the CREs emitting in
the radio frequencies.

Under ICCMB, it is expected that the non-thermal X-ray spectrum will have similar
spectral shape as of the radio spectrum if the IC X-ray and synchrotron
radio emission are produced due to the same population of electrons. This implies that the spectral
index in the radio should be the same as the slope of the X-ray spectrum.
In our observations, we find that the radio spectral index, $\alpha$ to be in the
range $1.4-2.5$ (see Figure~\ref{spind}). The X-ray emission photon
index ($\Gamma$) is related to the X-ray spectral index ($\alpha_{\rm Xray}$)
as $\alpha_{\rm Xray} = \Gamma - 1$. Our estimated value of
$\Gamma=1.86^{+0.49}_{-0.41}$ corresponds to $\alpha_{\rm Xray} =
0.86^{+0.23}_{-0.12}$, significantly flatter than that of the radio spectral
index estimated between the observed frequencies (see
Section~\ref{spec-index-map}).

In terms of energy, electrons emitting between 0.76 and 3.25 GHz in the
rest frame (i.e., 0.325 and 1.4 GHz in the observed frame), correspond to the
energy range $\sim$ $3.6-7.6$ GeV for magnetic field strengths of $3.3~\mu$G.
Thus, $\gamma_e$ of the electrons are $\sim7\times10^3$ and $\sim1.5\times10^4$,
respectively. For CMB photons at $z=1.325$, having temperature 6.34 K, the
Planck function peaks at $\nu_{\rm bg} \approx 6.6\times10^{11}$ Hz. These
photons are up-scattered by the CREs in the radio to X-ray frequencies
with average frequencies given by $\langle \nu \rangle \approx (4/3) \gamma^2
\nu_{\rm bg}$. Thus, the CREs emitting at rest frame 0.76 and 3.25 GHz
would up-scatter these CMB photons to $\sim4.3 \times 10^{19}$ and $\sim
2\times 10^{20}$ Hz, i.e., in the energy range $\sim$175 and 830 keV.
Therefore, our X-ray observations between $\sim0.3$ and 10 keV, corresponding to
$\sim0.7$ and 23.25 keV in the rest frame, do not probe this steep part of the
spectrum. In our observed range of X-ray energies, the emission arises due to
IC scattering from CREs at much lower energies ($\gamma_e\sim
5\times10^{2} - 1.8\times10^{3}$) corresponding to lower radio frequencies
where the spectral index is expected to be significantly flatter than that
between rest frame 0.76 and 3.25 GHz. In this way, the different spectral
indices as measured in the X-ray and radio observations can be explained.

\section{Magnetic field strength and age of the relic lobes}

\label{magfield-age}

\subsection{Magnetic field strengths}

\label{Magneticfield}

Together with radio observations, the ICCMB X-ray fluxes can be used to
constrain the magnetic field strength in the lobes of radio galaxies
\citep[e.g.,][]{erlund06}.  We use the formula given in \citet{tucker77} to
estimate the magnetic field in the lobes as,
\begin{equation}
\begin{aligned}
\frac{F_{c}}{F_{s}} &= 2.47 \times 10^{-19} \left(5.23 \times 10^{3}\right)^{\alpha} \left(\frac{T_{\rm CMB}}{\rm K}\right)^{3+\alpha}\\
& \,\, \times \frac{b(n)}{a(n)} \left(\frac{B}{\rm G}\right)^{-(\alpha +1)} \left(\frac{\nu_{c}}{\nu_{s}}\right)^{-\alpha}.
\end{aligned}
\end{equation}
Here, $F_c$ and $F_s$ are the flux of the X-ray and radio
emission, respectively, $T_{\rm CMB}$ is the temperature of the CMB at the
redshift of the source, $\nu_{c}$ is the frequency of IC X-ray
emission, $\nu_{s}$ is the frequency of radio synchrotron emission and the
constants $a(n)$ and $b(n)$ are taken from \citet{ginzburg65} and \citet{tucker77}, respectively.

This analysis requires the observed X-ray and radio fluxes to be co-spatial.  To
get the co-spatial flux, we extracted the source spectrum in a region matching
the 5$\sigma$ contours of the 0.325 GHz image, and fitted the spectrum with an
absorbed power-law model with $N_{\rm{H}}$ set to Galactic absorption. The best-fitting
photon index is found to be $\Gamma$ = 1.77$^{+0.51}_{-0.42}$ corresponding
to spectral index of 0.77$^{+0.22}_{-0.18}$. The co-spatial,  unabsorbed, 2 --- 10
keV rest frame X-ray flux is $1.44 \times 10^{-14}$ erg s$^{-1}$ cm$^{-2}$ that
corresponds to 0.75 nJy.
For a synchrotron flux density $F_s = 140$ mJy at
$\nu_s =0.325$ GHz, $\alpha = 0.77$ and $h\nu_{c} \approx 3$ keV, we estimate
the $B$ to be $\sim3.3~\mu$G.

\subsection{Energy equipartition}

We also compute the magnetic field strength assuming equipartition of energy
between CREs and magnetic field \citep[see][equation
2.77]{moffet75}. We use the 0.325 GHz emission, where the radio-emitting lobes
of the GRG are assumed to be cylindrical in shape, $\sim 500$ kpc long and
$\sim 180$ kpc in diameter each. The synchrotron cutoff frequencies are assumed to be
$\nu_{\rm min} = 10$ MHz and $\nu_{\rm max} = 10$ GHz.  We neglect the
contribution of relativistic protons to the total energy of relativistic
particles in our calculations. Note that, the 1.4 GHz flux density is heavily
affected by ICCMB losses. This analysis is valid only for synchrotron losses.
Therefore, the observed spectral index of 1.83 between 0.325 and 1.4 GHz 
would give rise to magnetic field strengths that are overestimated and would
also be sensitive to the choice of lower frequency cut-off. We therefore
assume a typical spectral index for synchrotron losses as 1.0. 
Using these values, we estimate the equipartition magnetic field strength,
$B_{\rm eq}\sim 3.5~\mu$G in the lobes.

The equipartition magnetic field strength and the magnetic field strength
derived using the X-ray and radio emission agrees well with each other
indicating the energy equipartition to be valid in the lobes of the GRG.

The energy equipartition can be independently tested by comparing the
total energies in the estimated magnetic field strength and the relativistic
electrons ($\mathcal{E}_{\rm e}$) \citep[see e.g.,][]{isobe11}. We find the
unabsorbed 2 --- 10 keV rest frame X-ray luminosity of the lobes to be
$1.64\times10^{44}~{\rm erg~s^{-1}}$, consistent with typical luminosities of
other GRGs (\citet[6C 0905+39;][]{erlund08}, \citet[3C 469.1;][]{laskar10}).
Using the X-ray luminosity, we estimate the minimum energy in the relativistic
electrons ($\mathcal{E}_{\rm e}$) following equation 4 in \citet{erlund06},
\begin{equation}
\mathcal{E}_{\rm e} = \frac{3}{4} \frac{L_{44}}{\gamma_{e} (1+z)^{4}} 10^{64}\ \textrm{erg}.
\end{equation}
Here, $L_{44}$ is the X-ray luminosity in units of $10^{44}\ \textrm{erg}\
\textrm{s}^{-1}$ and $\gamma_{e}$ is the typical Lorentz factor of the
electrons responsible for ICCMB. Assuming $\gamma_{e} = 10^{3}$, we find
$\mathcal{E}_{\rm e} = 4.2 \times 10^{59}$ erg. 
The total magnetic field energy ($\mathcal{E}_{B}$) in the lobes for
$B\sim3.3~\mu$G, within an cylindrical volume of $\sim180$ kpc diameter and
a total length of $\sim 1$ Mpc is $\sim6.4\times10^{59}$ erg. The two 
energies match within $\sim30$ per cent, indicating energy equipartition to be
valid. Similar equipartition of energy has been observed for the GRG DA 240 by \citet{isobe11}. Further, we note that the minimum energy in
the relativistic electrons for the GRG is similar to other powerful high
redshift Fanaroff$-$Riley type II radio galaxies \citep[see e.g.,][]{erlund06}.  It shows
that the ICCMB electrons that are longer lived than their radio emitting
synchrotron counterparts can deposit a significant amount of energy into their
surrounding IGM.

\subsection{Age of the relic}
\label{relicage}

The observed radio spectrum is well modelled by the JP model,
discussed in Section~3.3. In this case, $\nu_{\rm br}$ depends on the total age of the
electrons ($t_{s}$) and the magnetic field strength ($B$) as
\begin{equation}
\frac{\nu_{\rm br}}{\rm GHz} = 1.12 \times 10^{15} \left(\frac{B}{\rm G}\right)\left[\frac{2}{3}\left\{\left(\frac{B}{\rm G}\right)^2 + \left(\frac{B_{\rm IC}}{\rm G}\right)^2 \right\} \left(\frac{t}{\rm s}\right) \right]^{-2}.
\label{nubr}
\end{equation}

Thus, for our estimated value of magnetic field strength of $\sim3.3~\mu$G and
$\nu_{\rm br,rest}=1.32$ GHz, using Equation~\ref{nubr}, we find the age of the
relativistic electrons in the lobes and hence the age of the relic to be $t_{\rm s}$
$\sim8\times 10^6$ yr, which is typical for relic GRGs \citep{murgia11}.

\section{Summary} \label{summary}

We report the discovery of a relic GRG J021659-044920 at redshift ($z$) $\sim$
1.325 with extended diffuse X-ray emission, nearly co-spatial with the radio
lobes. The salient properties of this GRGs are as below.

\begin{enumerate}

\item The radio lobes are best detected at low radio
frequencies observed using the GMRT at 0.325 GHz.  The total angular extent at
0.325 GHz is 2.4 arcmin that corresponds to a projected linear size of
$\sim1.2$ Mpc at the redshift of the source.

\item The host galaxy is identified in deep optical (Subaru),
near-IR (UKIDSS) and mid-IR (SpUDS). It is a red ($R~-~z'$ =  2.0) and
dusty galaxy that brightens in mid-IR bands.  

\item The relic nature of the radio galaxy is evident as the AGN core, jets
and/or hotspots remain undetected and the lobes exhibits a very steep radio
spectral index, ${\alpha}_{\rm 0.325~GHz}^{\rm 1.4~GHz}$ $\sim$ 1.4 --- 2.5.
The 0.24---1.4 GHz radio spectrum of the lobe emission is
convex and steepens sharply above 0.325 GHz due to radiative losses.  The
spectral index of the lobes, estimated between 0.325 and 1.4 GHz, varies from
1.4 in the outer edges to 2.5 in the inner regions, suggesting backflow of
plasma.  

\item The X-ray spectrum between 0.3 and 10 keV is best fitted by
an absorbed power-law model with photon index, $\Gamma = 1.86_{-0.41}^{+0.49}$
for a fixed absorbing column density similar to the Galactic value. The
comparison of radio and X-ray spectral and morphological properties suggests
that X-ray emission is likely due to inverse-Comptonization of CMB photons by
low energy electrons compared to electrons radiating at 0.325 --- 1.4 GHz frequencies.

\item Using the ICCMB X-ray emission, we estimate the lower
limit for the total energy in relativistic electrons to be $\sim$ 4.2 $\times$
10$^{59}$ erg (for ${\gamma}_{\rm e}$ $\sim$ 10$^{3}$), implying
significant feedback from GRG into the surrounding IGM.

\item The magnetic field strength estimated using X-ray and radio emission and
by energy equipartition yield consistent field strengths of $\sim 3.5~\mu$G. This
suggests that the energy equipartition between magnetic field and relativistic
electrons to be valid in the GRG.

\end{enumerate}

\section*{Acknowledgements}

We gratefully acknowledge generous support from the Indo$-$French Center for the
Promotion of Advanced Research (Centre Franco- Indien pour la Promotion de la
Recherche Avanc{\'e}e) under programme no. 4404$-$3. We thank Sui Ann Mao, Rainer
Beck, Nirupam Roy and Dharam Vir Lal for critically reviewing the manuscript and giving
constructive and supportive suggestions. We thank Chris Simpson for providing
the 1.4-GHz VLA radio image of the SXDF field. We thank Alain Omont for useful
comments on an early draft of this manuscript. We thank the anonymous referee
for carefully reading the manuscript and giving useful comments and suggestions.
The GMRT is run by the National Centre for Radio Astrophysics of the Tata Institute
of Fundamental Research. The National Radio Astronomy Observatory is a facility
of the National Science Foundation operated under cooperative agreement by
Associated Universities, Inc. This paper is based in part on observations
obtained with {\it XMM$-$Newton} (an ESA science mission with instruments and
contributions directly funded by ESA Member States and the USA, NASA).  This
paper is partially based on data collected at Subaru Telescope, which is
operated by the National Astronomical Observatory of Japan. This work is
based in part on data obtained as part of the UKIRT Infrared Deep Sky Survey.
This work is based in part on observations made with the {\it Spitzer} Space
Telescope, which is operated by the Jet Propulsion Laboratory (JPL), California
Institute of Technology (Caltech), under a contract with NASA.



\bibliographystyle{mnras}
\bibliography{ref} 


\bsp	
\label{lastpage}
\end{document}